\documentclass[manuscript]{aastex}

\slugcomment{Submitted to Astrophysical Journal}
\shorttitle{Gamma Flares in Crab Nebula}
\shortauthors{Machabeli et al.}
\begin{document}
\title{On the origin and physics of Gamma Flares in Crab Nebula}
\author{George Machabeli, Andria Rogava and David Shapakidze}
\affil{Centre of Theoretical Astrophysics, Institute of Theoretical Physics, Ilia State University,
    Tbilisi, Georgia}

\begin{abstract}
We consider parametric generation of electrostatic waves in the
magnetosphere of the pulsar PSR0531. It is shown that in the
framework of this mechanism it is possible to convert the pulsar
rotational energy into the energy of Langmuir waves. The maximum
growth rate is achieved in the ``superluminal'' area, where phase velocity of
perturbations is exceeding the speed of light. Therefore
electromagnetic waves do not damp on particles. Instead, they create
plasmon condensate, which is carried out, outside of the pulsar
magnetosphere and reaches the Crab nebula. It is shown, that the transfer
of the energy of the plasmon condensate from the light cylinder to
the active region of the nebula happens practically without losses.
Unlike the plasma of the magnetosphere, the one of nebula contains ions,
i.e., it may sustain modulation instability, which leads to the
collapse of the Langmuir condensate. Langmuir wave collapse, in
turn, leads to the acceleration of the distribution function
particles. Furthermore, we consider processes leading to the
self-trapping of the synchrotron radiation, resulting in the growth
of the radiation intensity, which manifests itself observationally
as a {\it flare}. The condition for the self-trapping onset is derived, showing that
if the phenomena takes place at $~100 MeV$, then it doesn't happen 
at lower (or higher) energies. This specific kind of
{\it higher/lower energy cutoff} could explain why when we observe the
flare at 100 MeV no enhanced emission is observed at lower/higher
energies!
\end{abstract}


\section{Introduction}

Crab nebula is a source of almost steady high-energy emission. Observations made from orbital probes
(Fermi, SWIFT, RXTE) found an evidence of its variability in the X-ray range. Recently, Fermi and AGILE
satellites detected brief and strong bursts of gamma radiation above 100 MeV, with its source located
in the Crab nebula. One of such events was detected in September, 2010. Measurements made by AGILE on 12th
and 16th of April, 2011 have discovered new, unusually powerful bursts exceeding ``September'' ones
more than five times. Finally, in 2013, the most powerful bursts were observed \citep{al14}. In \citep{bb14}
an up-to-date review was given of the high-energy emission of Crab Nebula and the pulsar PSR 0532.

The one and only source of radiation energy in the nebula is the rotational energy damping of the pulsar
PSR 0532:

\begin{equation}
\frac{dE}{dt} = I \Omega \frac{d \Omega}{dt} = 5 \cdot 10^{38} \frac{erg}{sec}
\end{equation}

where $\Omega = 200 sec^{-1}$ is the angular frequency of the pulsar and $I = 10^{45} g \cdot cm^2$ is
its moment of inertia. This is more than enough to explain the value of the total luminosity
of the nebula $dW/dt = 5 \cdot 10^{38} erg/sec$. Note that the luminosity of the pulsar radiation
amounts only for about one percent of the total radiation intensity. In order to generate radiation with the
energy up to 100 MeV, it is necessary to have particles with the energy of the order of PeV's ($10^{15} eV$).
Tademaru \citep{ta73} and W.J. Cocke \citep{co75} noticed that  particles
injected by the pulsar in the nebula, due to the radiation losses, can not have energies exceeding 
TeV range ($10^{12}$). The possible
mechanism of the particle acceleration up to TeV energies and further production
of 100 MeV photons was suggested in \citep{cl12}, where gamma-bursts were explained in terms
of the Doppler enhancement due to the variation of the averaged mini jet orientations in space.
The latter, in its turn, happens due to the magnetic field line reconnection in the region occupied by
magnetically dominated plasma \citep{bf94, lu03}. It was also shown \citep{ki04} that in current layers,
where $E>B$, while sufficiently fast particles move along the electric field lines, particles are entrapped
by the layer and are accelerated. In \citep{cc12, cw12} a region was considered where the electric field
exceeds the normal component of magnetic field $E>B$. It was shown that the mechanism of particle
acceleration is similar to reconnection and for large-scale processes of the order of $10^{16} cm.$
particles are accelerated up to PeV energies, which corresponds to Lorentz factors of the order of
$3 \times 10^{9}$. An obvious test of the proposed mechanism could be a considerable change of the
structure of the magnetic field. So far such a noticeable change of this kind was {\it not} 
observed; in particular, Chandra failed to find such a variation in the X-ray range.

Let us consider the possibility of the energy transfer to the nebula by the kinetic energy of particles.
The mechanism of appearance of the particle acceleration in the pulsar magnetosphere was
outlined and specified in a number of papers: {\citep{kl54, de55, er66, gj69, st71, ta73, rs75, mi82}).
According to this mechanism maximum energy density of the electron flux is equal to $mc^2  \gamma_b n_{GJ}$,
where $n_{GJ}$ - is the number density of particles extracted from the pulsar surface by the
electric filed, generated by the neutron star rotation \citep{gj69}. Near the light cylinder $n_{GJ} =
10^6 cm^{-3}$. While $\gamma_b$ is the Lorentz factor of primordial particles. If we multiply this by the
volume of the light cylinder and divide by the time interval, necessary for the transfer of the energy
$R_{LC}/c$, we find out that the power provided by the particle flux is equal to:
\begin{equation}
mc^3 n_{GJ} \gamma_b R_{LC}^2 \approx 10^{33} erg/sec.
\end{equation}

Obviously, it is not sufficient. Therefore, it is necessary to have an additional mechanism of particle
acceleration. This mechanism will be proposed in the next section of the paper.
In the third section long wave-length electrostatic waves are
suggested and considered as candidates for the energy transport, transferring
energy from the pulsar to its nebula practically without any losses. 
In the fourth section we consider the process of particle
acceleration in the Crab nebula related to the collapse of Langmuir
waves \citep{za72}. The Langmuir wave collapse phenomenon was
widely discussed in the plasma theory (e.g., \citep{as79}), but it is relatively unknown to 
astrophysical community. In order to explain observed $100 MeV$ gamma bursts in the fifth section we
argue that their appearance is related to yet another plasma nonlinear
process - {\it self-trapping} - which could explain unusual properties of
these bursts. In the final, sixth, section, we summarize our results.  

\section{Parametric generation of Lengmuir waves in the pulsar magnetosphere}

Recently it was shown \citep{ma05, ma13} that relativistic centrifugal force can generate Langmuir waves.
According to Machabeli \& Rogava \citep{mr94} when a particle moves along a straight, rotating field line
the direction of centrifugal acceleration changes when particle's initial velocity  $v_0 < c/\sqrt{2}$ and
the particle decelerates. It was shown that this regime of the motion leads to the parametric growth of
Langmuir oscillations.

Let us write down the equation of motion in an inertial frame of reference:
\begin{equation}
\frac{\partial {\vec p}_i}{\partial t} + ({\vec v}_i \cdot \nabla)
{\vec p}_i = - c^2 \gamma_i ( 1 - \Omega^2 r^2)^{1/2} \nabla (1 -
\Omega^2 r^2)^{1/2} + q \left({\vec E} + \frac{1}{c} {\vec v} \times
{\vec B} \right)
\end{equation}
where $\vec p$, $\vec E$, $\vec B$ are momentum, electric field and magnetic field, respectively,
$\Omega$ is the angular velocity of rotation, $r$ - distance from the center of the pulsar to the particle,
$i=e^- , e^+$ - electron and/or positron. In the weak turbulence approximation, if we consider instantaneous
values of $\vec p, \vec E, \vec B$ as sums of their regular and perturbational components - $
\vec p = \vec p_0 + \vec p, ~  \vec E = \vec E_0 + \vec E', ~ \vec B = \vec B_0$  - then we find that:
\begin{equation}
\frac{\partial v_0}{\partial t} = \Omega^2 r \left( 1 - 
\frac{2 v_0^2}{c^2 (1 - \Omega^2 r^2/c^2)} \right)
\end{equation}
which yields the following simple, periodic solution for relativistic velocities:
\begin{equation}
v_0 = c \cdot cos(\Omega t + \phi)
\end{equation}
where $\phi$ is an initial phase and $c$ is the speed of light.

We can write also the continuity equation and the Poisson equation, taking into account that $n$ is the
fluctuation of number density - deviation of this quantity from its equilibrium value $n_0$, $v_0=p_0/m$
and $v=p/m$. This way we will obtain the closed set of equations for electrons and positrons. Further,
let us consider a model of two streams with different Lorenz factors. One of them we denote with the
index ``p'' and another with the index ``b''. These two streams are connected with each other by means of
the common electric field $E'$. Taking this fact into account and applying Fourier transform we reduce the
system of equations for both streams to the following equation \citep{ma13}:
\begin{equation}
\frac{(\omega^2 - \omega_p^2)}{\gamma_p^3}  N_p(\omega) = \frac{\omega_b^2}{\gamma_b} {\sum}_{l,s}
I_l(a) I_s(a) exp[i (l \phi_p - s \phi_b)] \left[ \frac{(\omega - (s-l) \Omega)^2}{(\omega -
s \Omega)^2} \right] N_p(\omega - s \Omega + l \Omega)
\end{equation}
where $N_p(t) = n_p exp[i\cdot a \textmd{sin}(\Omega t + \phi_p)]$,
$a \equiv kc/\Omega$, while $I_s(a)$ - are Bessel functions,
$s=1,2,3,...$ and the sum is taken from $-\infty$ to $+\infty$.
Taking the average of (6) by phases we see that on the right hand
side of the equation terms with $\phi_p = \phi_b$ give nonzero contribution when the harmonics
coincide $l=s$. The equality of phases means that particles at the
initial moment of time $t=0$ are situated at the equal distance from
the center of pulsar. The ultimate dispersion relation has the form:
\begin{equation}
\left( \omega^2 - \frac{\omega_p^2}{\gamma_p^3} \right) = \frac{\omega_b^2}{\gamma_b} \sum I_s^2(a)
\frac{\omega^2}{(\omega - s \Omega)^2}
\end{equation}
Taking $\omega = \omega_0 + \Delta$ we find out that $\omega_0^2 = 2 \omega_p^2/\gamma_p^3$
and $\omega_0= s \Omega $. The (7) relation contains two resonances which has to be satisfied simultaneously.
For $\Delta$ we obtain cubic equation, with its solution containing an imaginary part:
\begin{equation}
Im \Delta = \pm \sqrt{3}/2 (2 \omega_o \omega_b^2 I_s^2(a)/2)^{1/3}
\end{equation}
where $\omega_b^2 = 4 \pi e^2 n_b/m$, the Bessel function index  $s= \omega_o/\Omega$, $Im \Delta$ is
the growth rate - the quantity which specifies the growth rate of electrostatic perturbation. This
quantity depends on the quantitative value of the Bessel function. Its argument $a = kc/\Omega >>1$.
For large values of the argument the Bessel function has maximum value when the argument and index are
equal $s=a$ \citep{as64}.

Therefore, when $\omega_0 = k c$, in the parametric interaction of particles with waves participate only 
very high-order harmonics with $s>>1$. It is important to know what fraction of the Langmuir wave 
energy is dissipated in the pulsar magnetospheric plasma. An effective energy dissipation mechanism is 
Landau damping, which takes place when $v_{ph} < v$. For this purpose we must know at which wave numbers $k_c$
the dispersion curve crosses the $\omega = k c $ line. 

Electrostatic waves are described by the following equation:
\begin{equation}
1 - \sum \omega_p^2 \int \frac{f dp}{(\omega - k v)^2} \gamma^{-3} = 0
\end{equation}
Here the summation is made by the particle species - plasma electrons and positrons, $\omega_p^2 = 
8 \pi e^2 n_p/m$, while $f$ is the plasma distribution function normalized on unity. The (9) equation can be
easily solved in two extreme cases: $\omega >>kv$ and $\omega \approx kc$. The first case describes Langmuir waves, with
phase velocities is exceeding the speed of light and
\begin{equation}
\omega^2 = \omega_p^2 \gamma^{-3} + 3 k^2 c^2 
\end{equation}
For the second case in $(\omega - kv)^2$ we substitute $\omega \approx kc$ and $v = c(1- \gamma^
{-2})^{-1/2}$. If we expand the quadratic root by the small parameter $1/\gamma^2$ from (10) we 
derive the expression for the value of wave number vector $k_c$, at which the curve crosses the 
$\omega = k c $ line. 
\begin{equation}
\omega^2/ c^2 \equiv k_c^2 = 2 \gamma \omega_p^2/c^2 
\end{equation}
and the spectrum has the followng form \citep{lo79}:
\begin{equation}
\omega = kc[1 - \alpha (k - k_c)/k] 
\end{equation}

From the (8) condition it is evident that $\omega_0 < kc$, the $\omega_0 = kc$ condition is fulfilled
only when $k < k_c$. Therefore, the growth rate is maximum in the region where the phase velocity
 $\omega_0/k > \omega_0/k_c \simeq c$. But there are no particles with such velocities, hence Langmuir
 waves can interact with the particles of electron-positron plasma only via nonlinear effects.

 We have to note that while the parametric instability is developed the energy of the pulsar rotation
 is directly pumped into the energy of Langmuir waves. From (4) it is easy to estimate potential 
 power of the parametric instability:
\begin{equation}
\frac{dW}{dt} \simeq \frac{\omega^2 r^3 c^3 (m_e n_{LC})}{(c^2-\Omega^2 r^2)}  erg/sec
\end{equation}
While the longitudinal velocity $v_{\parallel}$ is decreased the linear velocity of rotation
$\Omega r$ is increased and consequently $W_t$ may reach the value
$5 \cdot 10^{38}$ at the light cylinder. Hence, we see that due to the parametric interaction, strongly
relativistic particles manage to pump the rotational energy of
the pulsar directly into the Langmuir waves.

Naturally, the question arises as to what causes the upper limit to the growth of the luminosity with
the value $5 \cdot 10^{38} erg/sec$. In our opinion the limit is related to the reconstruction of the
magnetic field from dipole to the monopole configuration \citep{mi82, ro03, os08, os09, bo01}.

We found out that the parametric generation of 
Langmuir waves for $\omega/k > c$ happens with higher growth rate
than for  $\omega/k \le c$. Therefore, Langmuir waves do not damp.
On the contrary, the Langmuir oscillations become accumulated and they transfer energy along
the magnetic field lines up to the Crab nebula. This process is considered in detail in the next chapter.

\section{Transfer of Langmuir oscillations to the nebula}

Let us make sure that the energy losses are negligible when the
waves are propagating towards the Nebula. It was assumed that the
most probable mechanism of longitudinal wave-particle interaction,
in the weak turbulence approximation, is a wave scattering by plasma
particles. This statement is true if the scattering process is due
to the resonance of the type of
\begin{equation}
(\omega_k-\omega_{k_1})/(k-k_1)=v 
\end{equation}
Here, if we take values of the frequencies from (10)
and (12), then it appears that the waves with the spectrum
$\omega\approx kc$ are scattered onto the thermal particles of
relativistic plasma, causing the energy pumping into the long-wave
band.

The plasma oscillations can be considered as a gas of quantum
quasi-particles - plasmons - (analog of photons) and the concept of
the number of plasmons is introduced. Scattering does not change the
number of plasmons, which is a conserved quantity:
\begin{equation}
N_k = |E_k |^2/\omega_k \simeq const 
\end{equation}
The energy is pumped to the long-wavelength region of $k\rightarrow
0$. From the conservation of $|E_k|^2/\omega_k$ it follows that
$|E_{k=0}|^2/|E_k|^2 \approx \omega_{k=0}/ \omega_k \approx
\gamma^{-2}$. Thus, the wave energy losses are quite significant.
This result has been obtained by \citep{lo79}, based on the assumption
that the waves are generated in the subluminal region, where
$v_{ph} < c$. 

However, in our case, the main bulk of energy is
generated in the region where $\omega_{k\rightarrow 0}$. Therefore,
the nonlinear scattering of Langmuir waves onto the plasma particles
should be considered in the region where $v_{ph} \gg c$.
Substituting the value of electrostatic wave frequency from equation
(10) in the resonant condition of nonlinear scattering, we obtain
the following expression:

\begin{equation}
v=\frac{\omega_p}{{\gamma_p}^{3/2}} \left(
\frac{1+3k^2c^2\gamma^3}{2{\omega_p}^2}-\frac{1+3k^{\prime 2}
c^2\gamma^3}{2{\omega_p}^2} \right)\frac{1}{k-k'}
\end{equation}

Taking into account that $k^2-k^{\prime 2}=(k-k')(k+k')$ and
assuming $k \sim k'$, we derive the following estimation:
$v/c\approx 3c/v_{ph}\ll 1$. The thermal particles of
this sort  in ultra-relativistic plasma practically do not exist and
therefore the scattering is not effective.

Since the density of the plasma particles is decreasing as the
wave propagates towards the Nebula, $\omega_{p^2}$ decreases as
well. However, $\omega_2$ should be conserved. In order for this to
happen, in the non-relativistic case, it is necessary the raise the
value of $k^2$. But in the relativistic case, the frequency
approximately equals $\omega_p \approx \omega_{LC} /
{\gamma_{LC}}^{3/2}$ when $k\rightarrow 0$. Therefore, it is sufficient to have:

\begin{equation}
\left(\frac{n_{LC}}{n_{17}}\right)^{1/2}\left(\frac{\gamma_{LC}}
{\gamma_p}\right)^{-3/2} \simeq 1
\end{equation}

Here, $n_{LC} \approx 10^{13}$ cm$^{-3}$ is the bulk plasma density
at the pulsar light cylinder; $n_{17} \sim 10^3$ cm$^{-3}$ denotes
the plasma density within the fibers of Crab Nebula \citep{mt77}. The
frequency does not change while Langmuir waves are propagating
towards the nebula. Therefore, as it follows from the condition
(17), the thermal particles of plasma have to be cooled down to
Lorentz factors of order $\sim 1-10$. It means that, due to the
ineffective wave-particle interaction, the energy of the long waves
$(k\rightarrow 0)$ is transferred to the distance, $R_{17}\sim
10^{17}$ cm, almost without any considerable losses.

In the Nebula, the spectrum of the Langmuir oscillations is also
situated beyond the light cylinder $k\rightarrow 0$. The energy of
electrostatic oscillations has to be converted into the kinetic
energy of the particles. This issue is considered in the following
section.

\section{Collapse of Langmuir oscillations}

``Condensate'' -is an intense gas of long-wave plasmons which is
unstable when the background is modulated by low-frequency waves.
But there are no such waves in the pulsar magnetosphere, since the
magnetosphere consists exclusively of electrons and positrons.
In the Crab nebula, unlike the magnetosphere, there are ions and
they can ensure the low-frequency modulation of the plasmon
background, which, in its turn, leads to modulational instability
\citep{vr62}. The development of this instability manifests itself
by the formation of caverns - regions of localization of excessive
wave energy. Due to this instability the energy of Langmuir waves is
localized in the cavern. Plasmons appear to be locked in the cavern.
High-frequency pressure pushes particles out of the cavern.
Consequently the cavern deepens and it absorbs in itself more and more
plasmons. The process of spatial attraction of plasmons is
accelerated. It is accompanied by the collapse of the cavern. 

This phenomenon can not be considered in the framework of the weak
turbulence approximation. The dynamics of cavern collapse was studied
in \citep{ga77}. A numerical experiment,
illustrating the dynamics of the cavern collapse was presented in
\citep{de76}. It turned out that in the process of cavern collapse
the wave vector $\vec k$ grows until the collapse phase velocity
equals the speed of particles $\omega/k = v$. Particles which fall
in the resonance with the wave extract the wave energy. Due to Landau
damping the cavern collapse stops to develop and the distribution function of
the particles acquire a long tail \citep{pe82}.

The above-described collapse scenario takes place in the plasma without external magnetic field. 
In Crab
Nebula plasma number density is $n_0 = 10^3$ and magnetic field $B_0=10^{-3} G$, implying that 
$\omega_p >>
\omega_B \equiv eB_0/mc$ and the influence of the magnetic field on the collapse of Langmuir waves
can be neglected. In this case the equation for the amplitude of Langmuir oscillations can be
written in the following way:
\begin{equation}
\frac{\partial^2 E}{\partial t^2} -
\frac{\omega_p^2}{\gamma_p^{3/2}} E - \frac{3 r_D^2
\omega_p^2}{\gamma_p^{3/2}} \frac{\partial^2 E}{\partial x^2} =
\frac{\omega_p^2}{\gamma_p^{3/2}} \left(\frac{n}{n_0}\right) E
\end{equation}
At the other hand the number density perturbation, resulting from the plasmon pressure, obeys the
equation:
\begin{equation}
\frac{\partial^2 n}{\partial t^2} - (T_e/M) \frac{\partial^2
n}{\partial x^2} = \frac{\pi M}{16} \frac{\partial^2}{\partial x^2}
|E|^2,
\end{equation}
where $M$ is the ion mass. Auto-model solution of the system have the ``burst-like'' 
appearance:
\begin{equation}
|E|^2 = \frac{E_0^2}{(\tau_0 - \tau)^2}
\end{equation}
where $\tau \equiv n_0 T_e(\omega_p/\gamma_o^{3/2})t E_0^2/8 \pi$.

Here $|E_0|^2$ is the field energy density at the initial stage of
the modulational instability and $\tau_0$ - is the time of the active
phase of the collapse \citep{za72}. In the course of the collapse,
when waves are being damped, electric field is decreasing with the
increase of the particle kinetic energy.

\section{Synchrotron radiation of the nebula and ~ 100 MeV bursts}

As early as in 1953 I.S. Shklovsky in order to explain the
continuous radio emission of the Crab Nebula suggested the synchrotron
mechanism \citep{sh53}. Soon afterwards observations in the visible
spectrum range revealed a substantial level of linear polarization
\citep{va54, do54, ow56}. The variation of the polarization degree
in various regions evidently indicates that synchrotron mechanism
has indeed the leading role in the generation of the nebula radiation.

In \citep{go65, we89, ja96} the mechanism of Compton-synchrotron
formation of spectra was proposed. Above a few hundreds of MeV the
intensity of synchrotron emission is strongly diminished and starting
from frequencies of the order of 1 GeV the radiation is determined by
the inverse Compton scattering. In this paper we will restrict
ourselves by considering frequencies of the order of $100 Mev$,
which means we consider only the synchrotron mechanism of the
radiation. However, we have to note that the problem of the pulsar
rotational energy transfer to the Nebula is open and actual also for
the inverse Compton radiation. Not rejecting the idea of explanation
of gamma bursts in terms of the magnetic field line reconnection,
which was mentioned in the Introduction, we would like to
consider a number of nonlinear phenomena, related with the influence
of the powerful radiation on the medium \citep{as62, wh74, ak67,
ch64, kl66}. As it turns out, bursts with $~100MeV$ energies are not
accompanied by any tangible increase of radiation intensity in any
other ranges of the Crab Nebula radiation \citep{al14}. Such
behavior of the bursts makes us surmise that this phenomenon can be
explained by the self-trapping effect. In this case the mysterious
behavior of the bursts could be explained in a quite natural way.

Self-trapping is a well studied phenomenon in nonlinear optics, its
theory was developed in 1960s, accompanying the
appearance of powerful light sources - lasers. As it was mentioned
above in Crab Nebula a modulational instability is developed which
leads to the appearance of caverns, where the energy of long
wavelength electrostatic oscillations is stored. Caverns are
eventually collapsed. At the beginning of the process the collapse
time scale is determined by the growth rate of the modulational
instability. Further, while $\tau \rightarrow \tau_0$, according to
(20), the collapse rate increases in a burst-like way; the active
phase of the collapse starts and Landau damping makes substantial
the absorbtion of the energy by cavern particles from Langmuir
waves. Particles get accelerated up to required limits and the cavern
collapse stops. The size of the cavern throughout its active phase
is estimated as a few  Debye radii wide \citep{as79}. As it is
well-known, Debye radius is the scale of species separation in plasma
and is determined  by the length scale, over which electron density
perturbation may be shifted  due to its thermal motion over the
period of plasma oscillations. In our, relativistic, case:
\begin{equation}
r_D^2 = \gamma_p^3 c^2/ \omega_p^2
\end{equation}
For the bulk of the plasma particles in the Crab Nebula the particle number density is of the order of
$n_0 \approx 10^3$ \citep{mt77}. Prior to acceleration Lorentz factors of plasma particles are
of the order of $\gamma_p \approx 1-10$ \citep{mu89}. After the acceleration happens, the energy 
maximum, as it follows from
observations, falls in the soft X-ray range. That is why one can assume that the majority of plasma
particles are accelerated up to Lorentz factors $\gamma \approx 5 \cdot 10^5$. Then the frequency
of synchrotron radiation is situated in the right range:
\begin{equation}
\omega_s = (eB/mc) \gamma^2 = 1.8 KeV
\end{equation}

Putting the values of the number density and Lorentz factor in (21) we obtain that $r_D = 5 \times
10^{14} cm$. An average distance between plasma particles is $r_{cp} \approx n^{-1/3} = 10^{-1}$, hence
we can estimate maximum quantity of particles in Debye volume as $N_D \approx (r_D/r_{cp})^3 \approx
10^{46}$.

In the process of the cavern collapse participate also particles from the 
distribution function tail,
which are being accelerated up to Lorentz factors $\gamma \approx 3 \times 10^9$. 
From (22) we can estimate
that particles with
such Lorentz factors are emitting photons with energies about $100 MeV$ due to the synchrotron 
radiation
mechanism. Assuming that $n_p \gamma_p = n_s \gamma_{sy}$ we can find the number of particles $N_s
\approx 10^{38}$. It is close to the value found in observations \citep{al14}. Multiplying the
number of particles on the radiation intensity for one particle:
\begin{equation}
\left(\frac{d \varepsilon}{dt}\right)_s  = \left(\frac{2 e^2 \omega_B^2}{3c}\right) \gamma^2
\end{equation}
we obtain the intensity of the synchrotron radiation of the order of $100 MeV$.
\begin{equation}
\left(\frac{\partial \varepsilon}{\partial t} \right)_s \approx 10^{36} erg/s
\end{equation}

The electromagnetic wavelength can not be less than $r_D$ since in the Debye volume electromagnetic
field is screened out due to the grouping of charged particles, that is, the polarization of the 
medium takes
place. For convenience, let us imagine the following model: the medium within the Debye radius consists
of dipoles with the same spatial orientation. In this case the polarization vector:
\begin{equation}
\vec P = N(e \vec r)
\end{equation}

In the course of the collapse transverse size of the cavern (in the X and Y directions) is decreasing.
The acceleration of electrons happens in the same directions, which, in turn, generates synchrotron
photons. Let us assume that local magnetic field is directed along the Z axis and 
the observer is located
in the X direction. Synchrotron radiation is directed in the X and Y direction. The electric field of
the radiation $E(t)$ shifts charged particles. The shifting causes the appearance of the elastic 
counterforce $\vec f(t) = - \eta \vec r(t)$, where $\eta$ is the elasticity coefficient. For not 
too small values of $E(t)$ the elasticity force has nonlinear form:
\begin{equation}
\vec f(t) = - \eta \vec r(t)  - q {\vec r(t)}^3
\end{equation}

The value of the electron shifting is determined from the equation of motion. In the relativistic case
it has the following form:
\begin{equation}
m \gamma^3 \frac{d^2 \vec r}{dt^2} - m \Gamma \frac{d \vec r}{dt}  + \eta \vec r + q {\vec r}^3 =
e \vec E
\end{equation}
where the second term on the left hand side defines the dissipation - damping rate. Taking into account
the definition of the polarization vector (Eq. (25)) and $\eta = m \gamma \omega_0^2$ we obtain:
\begin{equation}
\frac{d^2 \vec P}{dt^2} + \Gamma \frac{d \vec P}{dt}  + (\omega_0^2 / \gamma^2)  \vec P
+ (q/me^2 N^2 \gamma^3) \vec P^3 =
e^2 N {\vec E} /m \gamma^3
\end{equation}

In our case $\omega_0$ is the frequency of Langmuir
oscillations $\omega_o = \omega_L/\gamma^{3/2}$. The $\vec E$ field
is large but still it is much less than the internal field of the
cavern. In particular, the power of the wave energy of the cavern
stored in Langmuir waves is of the order of $10^{38} erg/s$, whereas
the synchrotron radiation intensity is of the order of $10^{36}
erg/s$. Consequently, the electric field of the wave is small
compared to the internal field. In such circumstances the nonlinear
term can be considered small and the equation can be solved by the
method of consecutive approximations. Splitting $\vec P = \vec
P_{L} + \vec P_{NL}$, with $\vec P_{L}
>> \vec P_{NL}$ and neglecting the nonlinear term we get:
\begin{equation}
\frac{d^2 \vec P}{dt^2} + \Gamma \frac{d \vec P}{dt} + (\omega_0^2 / \gamma^2)  \vec P  =
e^2 N {\vec E} /m \gamma^3
\end{equation}
Taking $\vec E = \vec A cos(\omega t)$ we can find the solution:
\begin{equation}
\vec P(t) = e^2 N \vec A cos (\omega t + \Phi)/m \gamma^3 \sqrt{(\omega^2 - \omega_0^2)^2 + 4 \Gamma^2
\omega^2}
\end{equation}
where $tg \Phi = \Gamma \omega/(\omega^2 - \omega_0^2)$.

Note that the Langmuir frequency is much smaller than the synchrotron radiation
frequency $\omega_0 << \omega$. Therefore, the range of frequencies which we are considering is
far from the resonance: $|\omega_0^2 - \omega^2|>> 4 \Gamma^2$. 
It means that the contribution related
to the dissipation can be neglected. The polarization vector $\vec P$ is related to the electric field
via the polarizability of the medium $\mu$, in the following way: $\vec P = \mu \vec E$. This 
expression
can be written also as:
\begin{equation}
\vec P(t) = \mu(\omega) \vec E(t)
\end{equation}

The nonlinear approximation equation has the following form:
\begin{equation}
\frac{d^2 \vec P_{NL}}{dt^2} + \omega_0^2 \vec P_{NL} = - q \mu^3(\omega) 
{\vec E(t)}^3/m \gamma^3 e^2 N^2
\end{equation}

Let us rewrite ${\vec E^3(t)}$ using the trigonometric identity: $cos^3(\omega t) = 
(1/4)( cos \omega t +
cos 3 \omega t)$. This way on the right hand side of (32) we have two terms 
describing input of the first
and the third harmonics. Consequently we can write:
\begin{equation}
\vec P(t) = \mu(\omega, A) \vec E(t)
\end{equation}
\begin{equation}
\mu(\omega, A) = \mu(\omega) [ 1 + 3 q \mu^2(\omega) A^2 /4mn^2 e^2 \omega^2]
\end{equation}
and $\mu(\omega)$ is determined from the solution of (29):
\begin{equation}
\mu(\omega) = e^2 N/m\gamma^3 \omega^2
\end{equation}
Note that in the series expansion of $\mu(\omega, A)$ we retain first non-vanishing terms.

The dielectric permittivity of the medium is described by the tensor $\varepsilon_{ij}(\omega, \vec E)$.
The connection between this tensor and $\mu_{ij}(\omega, A)$ is determined by the expression:
\begin{equation}
\varepsilon_{ij}(\omega, \vec E) = \delta_{ij} + 4 \pi \mu_{ij}(\omega, A)
\end{equation}

The induction vector $\vec D = \vec E + 4 \pi \vec P$, $D_i =
\varepsilon_{ij}(\omega, \vec E) E_j$. After this is written down,
taking into account (36) and (34), we write down Maxwell equation:
\begin{equation}
\nabla \times \vec B = (1/c)[ \varepsilon(\omega) + 3 \pi q \mu^3(\omega) A^2/mN^2 e^2 \omega^2] \frac{\partial \vec E}{\partial t}
\end{equation}

From this analysis it is clear that the influence of the nonlinear
term is equivalent of the change of the dielectric permittivity or
the refraction index of the medium. When an electromagnetic wave is
propagating in the medium the refraction index $H = c/v_{ph}$ is a
function of frequency. Therefore, the dispersion of light depends on
the refraction index. From the expression $H^2 = \varepsilon$ we
find that in the cavern, which undergoes the active phase, the
refraction index becomes equal to $H = H_L + H_{NL}$, where
\begin{equation}
H_L^2 = \varepsilon(\omega), ~~ H_{NL} = H_2 A^2, H_2 = 6 \pi q \mu^3(\omega)/mN^2e^2 \omega^2
\end{equation}

Therefore, if $H_2>0$, the refraction index in the cavern $H = H_L +
H_{NL}$ turns out to be larger than the refraction index beyond the
cavern, which is equal to $H=H_L$. In the whole volume of the cavern
let us separate rays which are directed towards the observer. These
rays are generated by particles with Lorentz factors $\gamma = 3
\times 10^9$ and directed towards the observer. That's why they are
emitted within the narrow cone with the opening angle $1/\gamma$.
Due to the linear diffraction they have to diverge - diffuse  in the
direction normal to the propagation direction and before leaving the
cavern they will be located within the cone with opening angle $2
\theta_D$, where $\theta_D \approx \lambda/ r_D H_L$. However, when
these rays leave the nonlinear medium of the cavern and enter the
space with refraction index $H_L$ they undergo nonlinear
refraction. If a ray falls on the boundary between nonlinear,
optically thicker medium and linear, optically thinner medium and
if the incident angle satisfies the condition $\theta_0 > \theta_D$
then all diffracted rays will undergo complete internal reflection.
We are interested in the condition $\theta_0 \approx \theta_D$ which
leads to the situation when the rays are combined to form a parallel
beam and the observer witnesses an increase of the radiation
intensity.

An asymptotic sliding angle for the complete reflection from the cavern boundary is determined
in the following way: $cos \theta_0 = H_{L}/(H_{L} + H_2 A^2)$. For the small $\theta$ angle
we find:
\begin{equation}
\theta_0 \approx 2 (H_2/H_{L}) A^2
\end{equation}
From the condition $\theta_0 \approx \theta_D$ we determine the length of those 
waves which are focused
in a parallel beam:
\begin{equation}
\lambda \approx 2 A^2 H_2 r_D
\end{equation}

Substituting (35) and (38) we find out that $H_2 \simeq 1/\omega^8$
and if the self-trapping condition is satisfied for frequencies
close to $100MeV$, from the observed dependence of the radiation
intensity on frequency (see, e.g., Fig. 1 in \citep{ja96}) or Fig. 2
in \citep{bb14}) we conclude that for other frequencies the (37)
condition is not met. This circumstance could explain strange
behavior of bursts in the $100MeV$ region of the spectrum, while at
the same time in other frequency ranges the increase of the
intensity is not observed.

\section{Conclusion}

In the conclusion we just briefly summarize the contents and main results of this paper:

\begin{itemize}
\item
The possibility of the direct pumping of energy from the neutron star's vast 
rotational kinetic energy storage
$5 \cdot 10^{38} erg/sec$ to proper electrostatic plasma (Langmuir) oscillations is demonstrated.
\item
It is shown that the growth rate of the perturbations is maximum in the ``superluminal'' area, 
where phase velocity of
perturbations is exceeding the speed of light. That is why in this
region the condensate of plasmons is formed, which is transferred to
the Crab Nebula.
\item
It is demonstrated that the transfer of the energy of the plasmon condensate from the pulsar
magnetosphere to the nebula over the huge distance $~ 3 \cdot 10^{17} cm$ takes place practically
without any tangibles losses.
\item
Unlike the pulsar magnetosphere in the nebula apart from electrons 
and positrons there are also protons.
That is why a modulational instability is developed, which leads to the collapse 
of Langmuir waves. A cavern
is formed which collapses and on the final stage of the collapse particles 
attain very high Lorentz
factors, resulting in the powerful emission of the nebula and the collapse stops at the scale of a few
Debye radii.
\item
Theoretical estimation of the cavern (radiating region) and the
number of emitting particles coincides with evaluations made on the
basis of observational results in \citep{al14}.
\item
It is shown that in the course of the active phase of the collapse
in the cavern, due to the influence of nonlinear processes on the
polarization properties of the medium, self-trapping of the
synchrotron radiation generated within the cavern, takes place.
\item
It is shown that if the conditions for the appearance of
self-trapping phenomenon are fulfilled for certain values of emitted
wave frequencies, for other, both higher and lower values of the
frequency, they are not satisfied.
\end{itemize}

\end{document}